\documentclass[aps,prd,reprint,showpacs,superscriptaddress,twocolumn]{revtex4-2}
\usepackage{amsfonts}
\usepackage{amsmath}
\usepackage{amssymb}
\usepackage{dsfont}
\usepackage{hyperref}
\usepackage{graphicx}
\usepackage{subcaption}
\usepackage{mathtools}
\usepackage{float}
\usepackage[usenames,dvipsnames]{xcolor}
\usepackage[normalem]{ulem}
\usepackage{times}
\usepackage{braket}
\usepackage{booktabs}
\usepackage{multirow}
\usepackage{tikz}
\usepackage{placeins}
\usepackage[font=normal, justification=Justified]{caption}

\hypersetup{    colorlinks=true,linkcolor=blue,citecolor=blue,filecolor=blue,urlcolor=blue,breaklinks=true}
\RequirePackage{color}

\begin{document}

\title{Dynamics of massive and massless particles in the spacetime of a wiggly cosmic dislocation}

\author{Frankbelson dos S. Azevedo}
\email{frfisico@gmail.com}
\affiliation{Departamento de F\'{\i}sica, Universidade Federal do Maranh\~{a}o, 65085-580 S\~{a}o Lu\'{\i}s, Maranh\~{a}o, Brazil}

\author{Edilberto O. Silva}
\email{edilberto.silva@ufma.br}
\affiliation{Departamento de F\'{\i}sica, Universidade Federal do Maranh\~{a}o, 65085-580 S\~{a}o Lu\'{\i}s, Maranh\~{a}o, Brazil}

\date{\today}

\begin{abstract}
In this paper, we investigate the spacetime containing both small-scale structures (wiggles) and spatial dislocation, forming a wiggly cosmic dislocation. We study the combined effects of these features on the dynamics of massive and massless particles. Our results show that while wiggles alone lead to bound states and dislocation introduces angular momentum corrections, their coupling produces more complex effects, influencing both particle motion and wave propagation. Notably, this coupling significantly modifies radial solutions and eigenvalues, with the direction of motion or propagation becoming a critical factor in determining the outcomes. Numerical solutions reveal detailed aspects of particle dynamics as functions of dislocation and string parameters, including plots of trajectories, radial probability densities, and energy levels. These findings deepen our understanding of how a wiggly cosmic dislocation shapes particle dynamics, suggesting new directions for theoretical exploration in cosmological models.
\end{abstract}
\maketitle

\section{Introduction}

Cosmic strings are topological defects that may have formed during the expansion and cooling down of the Universe through mechanisms of spontaneous symmetry breaking. Although their detection nowadays is still an open subject of study, their effects on the Universe's early evolution are well established \cite{TWBKibble_1976,ade2014planck}.
These gravitational objects have a mass distribution often characterized as straight, with infinite length and very thin thickness. However, a more accurate description of their core includes substantial small-scale structures, such as wiggles, on scales much smaller than their correlation length. Consequently, a ``wiggly'' string has a gravitational field distinct from that of an ideal string without wiggles. Due to these small-scale structures, a non-vanishing Newtonian potential is produced, similar to that found in a massive rod \cite{Vilenkin1994,PhysRevD.41.3038}.
The presence of wiggles on cosmic strings can lead to intriguing effects in astrophysical contexts. For example, when two particles move on opposite sides of a wiggly string, they experience a velocity difference partially due to the gravitational attraction induced by the wiggles \cite{PhysRevD.45.1884,PhysRevLett.67.1057}. Furthermore, the wiggles can enhance the efficiency of forming large-scale structures and lead to the generation of a primordial magnetic field in the Universe \cite{PhysRevD.45.3487,VACHASPATI1993355,PhysRevLett.67.1057}. These wiggles are also responsible for the accretion of dark energy \cite{doi:10.1142/S0218271806008322}.

Cosmic strings with small-scale structures along their length, often referred to as wiggly cosmic strings, have their spacetime metric expressed in the following form \cite{PhysRevD.41.3038, Vilenkin1994}:
\begin{equation}
ds^{2} = -N(r)dt^{2} + dr^{2} + L(r)d\theta^{2} + M(r)dz^{2}, \label{metric}
\end{equation}
where the coefficients of the metric are $N(r)= 1 +  \mathnormal{w} \ln\frac{r}{r_0}, 
M(r)= 1 -  \mathnormal{w} \ln \frac{r}{r_0} ~ \text{and} ~ 
L(r)=\alpha^2 r^2$ (this metric set $c = 1$).
The term $\alpha=1-4G(\tilde{\mu}+\tilde{T})$ accounts for the conical geometry of the spacetime, and $ \mathnormal{w}=4G(\tilde{\mu }-\tilde{T})$ accounts for the excess energy density, indicating the presence of wiggles. To be precise,
the terms $\tilde{\mu}$ and $\tilde{T}$ represent the effective mass per unit length (energy density) and tension on the string, respectively.
They are related through the equation of state $\tilde{\mu}\tilde{T}=\mu^2=\text{const.}$, where $\mu$ is the unperturbed string energy density. The term $G(\tilde{\mu}-\tilde{T})\approx 10^{-6}$ \cite{MBHindmarsh_1995}, which makes the logarithmic term a small perturbation. Therefore, the propagation of particles is only valid within the limit $r_0 < r \ll r_0 e^{1/ \mathnormal{w}}$, with $r_0$ being a small number that denotes the effective string radius. For Grand Unification Theory (GUT) scale strings, the value is $r_0 \approx 10^{-30}$ cm \cite{PhysRevD.42.2669}. Anyway, the upper limit in this inequality is still very large, making our calculation valid for extremely large distances from the defect. This fact was also observed in Ref. \cite{PhysRevD.45.1884}.

It is well known that certain cosmic string solutions predict the existence of spinning strings, where the strings themselves rotate \cite{EPJC.2019.79.311,AoP.2015.356.346,PhysRevD.70.047502,PRD.2020.102.105020,AHEP.2021.2021.6709140,Universe.2020.6.110203}. An observer moving with such a spinning cosmic string would perceive spacetime as twisted, exhibiting a helical structure along the string's direction. This metric spacetime with these two factors (spin and twist) is often called a chiral cosmic string \cite{10.1063/1.531995,PhysRevD.47.4273}.
However, if we neglect the rotation of this spacetime, the resulting geometry resembles that of what is named a cosmic dislocation. This geometric effect can be understood through a cut-and-glue process: cutting out a wedge of a four-dimensional spacetime, displacing one face vertically, and then gluing them back together \cite{PhysRevD.47.4273,PhysRevD.70.047502,KPTod_1994}. In the past and still today, the physical implications of the effects caused by this spatial dislocation defect have been studied from various perspectives, in both material media \cite{DEPADUA1998153,C.Furtado_2000,PLA.2001.288.33,PLA.2015.379.2110,PLA.2016.380.3847,AoP.2021.433.168598,EPJP.2019.134.131} and the cosmological context \cite{doi:10.1142/S0219887823500676,Ma_2016,IJMPA.2020.35.2050195,EPJP.2020.135.691,CJP.2020.66.587,AHEP.2020.2020.4832010,EPL.2020.132.50006,GRG.2022.54.69,FBS.2024.65.6}.

In this work, we propose that the string described by the metric \eqref{metric} spins off, twisting the wiggly string and resulting in a {\it wiggly cosmic dislocation}. Thus, a cosmic string spacetime with both small-scale structures and dislocation effects can be described by
\begin{align}
ds^{2} = -N(r)dt^{2} &+ dr^{2} + L(r)d\theta^{2}  \notag \\
 &+ \left( \sqrt{M(r)}dz + \chi d\theta  \right)^{2}, \label{metric2}
\end{align}
where $0 \leq \chi < 1$ \cite{EPJP.2019.134.131,doi:10.1142/S0219887823500676} is the dislocation parameter. This parameter is analogous to the Burgers vector of a screw dislocation in solid continua \cite{RolandAPuntigam_1997}. By setting $\chi=0$ in Eq. \eqref{metric2}, we recover the wiggly cosmic string spacetime given by Eq. (\ref{metric}). Also, if we set $\tilde{\mu }=\tilde{T}$ ($ \mathnormal{w}=0$) in Eq. \eqref{metric2}, we recover the straight cosmic string with dislocation, as expected \cite{PhysRevD.71.024005}. 

We choose not to consider, in the spacetime \eqref{metric2}, the spin parameter due to the rotation of the string for two reasons: first, we aim to simplify our future calculations. Second, the metric spacetime with only the dislocation parameter is often studied on its own. Additionally, when the spin parameter is smaller than the dislocation parameter, a Lorentz transformation exists such that the background of a chiral cosmic string can be viewed as that of a cosmic dislocation (the corresponding spacetime is static) \cite{AoP.2022.436.168666,PhysRevD.47.4273, PhysRevD.70.047502}.

Some effects of the small-scale structures and dislocation are known separately. However, in the metric \eqref{metric2}, these parameters (energy density of the wiggly string and dislocation parameter) are coupled, which can give rise to novel effects. In the following sections, we will investigate these new effects on particle motion by studying the geodesics and, subsequently, explore their impact on wave propagation by analyzing the wave functions.

\section{Geodesics in the Spacetime of a Wiggly Cosmic Dislocation}
\label{section1}

In this section, we investigate the effects of a wiggly cosmic dislocation, characterized by the coupling between small-scale structures and dislocation, on the geodesic equation for massive and massless particles. The derivation of the results presented below follows the same Lagrangian approach as in Ref. \cite{bartelmann2019general}, where the geodesic equations are solved for the effective Lagrangian in the Schwarzschild spacetime within the equatorial plane.

Let's write the Lagrangian for a particle in the wiggly cosmic dislocation 
spacetime as
\begin{align}
 \mathcal{L} = \frac{1}{2}g_{\mu \nu }\dot{{x}}^{\mu } \dot{x}^{\nu } 
= \frac{1}{2}&\left(-N(r)\dot{t}^{2} + \dot{r}^{2} + (L(r)\chi^2)\dot{\theta }^{2} \right. \notag \\
&\left. + \,2\chi\sqrt{M(r)}\dot{\theta}\dot{z} + M(r)\dot{z}^{2}\right), 
\label{Lagrangian}
\end{align}
where ``$\cdot{}$'' indicates a derivative concerning an affine parameter (the proper time). From Lagrangian (\ref{Lagrangian}), we write the Euler-Lagrange equation for each component to find the constants of motion 
\begin{align}
&K= -N(r)\dot{t}, \notag \\ 
&J= ( L(r)+ \chi^2)\dot{\theta}+\chi\sqrt{M(r)}\dot{z}, \notag  \\ 
&Z=M(r)\dot{z}+\chi\sqrt{M(r)}\dot{\theta} . \label{CteMotion}
\end{align}
We can recognize $K$ as the energy, $J$ as the angular momentum aligned with the string's axis ($z$-axis), and $Z$ as the linear momentum along $z$-direction. The dislocation parameter $\chi$ and the wiggly parameter $\mathnormal{w}$ affect both angular and linear momentum.

From Eq. \eqref{Lagrangian}, we use $2 \mathcal{L}=\epsilon $ ($\epsilon=-1$ for massive and $\epsilon =0$ for massless particles) and manipulate the constants of motion \eqref{CteMotion} to find
\begin{eqnarray}
\dot{r}^{2}+\frac{\Omega(r)^{2}}{L(r)}+\frac{Z^{2}}{M(r)}-\frac{K^{2}}{
N(r)}-\epsilon  &=&0, \label{Equationoftotalenergy}
\end{eqnarray}
with $\Omega(r)=J-\frac{\chi Z}{\sqrt{M(r)}}$
being the angular momentum constant plus a term that carries a couple of dislocation with wiggles. Nevertheless, this couple only exists when the linear momentum along the string's axis is non-null.  

Since our primary goal is to find the orbit $r(\theta )$, we use $r{^{\prime}}\left( \theta \right) =\frac{dr}{d\theta }=\frac{\dot{r}}{\dot{\theta }}$ to transform Eq. \eqref{Equationoftotalenergy} to
\begin{eqnarray}
r^{\prime 2}+L(r)+\frac{L(r)^{2}}{\Omega(r)^{2}}\left(\frac{Z^{2}}{
M(r)}-\frac{K^{2}}{N(r)}-\epsilon \right) &=&0. \label{termsr}
\end{eqnarray}
Now, we make the change of variable $u\equiv 1/r$ and $u^{\prime }=-r^{\prime }/r^{2}=-u^{2}r^{\prime }$ to find
\begin{eqnarray}
u^{\prime 2}+u^{4}L(u)+\frac{u^{4}L(u)^{2}}{\Omega(u)^{2}}\left( \frac{Z^{2}}{M(u)}-
\frac{K^{2}}{N(u)}-\epsilon \right) &=&0, \label{termsu}
\end{eqnarray}
where 
$
N(u)= 1 + \mathnormal{w} \ln \frac{u_0}{u},
M(u)= 1 - \mathnormal{w} \ln \frac{u_0}{u} 
$
\text{and}
$L(u)=\alpha^2/ u^2$
are the coefficients of the metric in terms of $u$ and 
$
\Omega(u)=J-\frac{\chi Z}{\sqrt{M(u)}}.
$

Finally, the differentiation of Eq. \eqref{termsu} concerning $\theta$ yields the trivial solution $u^{\prime}=0$, which implies a
circular orbit; and in addition, the left orbit equation to be solved is
\begin{align}
u^{\prime \prime } + u \alpha^2 -  &\frac{ \mathnormal{w} \alpha^4}{2u\Omega(u)^2} \left[ \frac{Z^2}{M(u)^2} + \frac{K^2}{N(u)^2} + \frac{\chi Z}{M(u)^{3/2} \Omega(u)} \right. \notag \\
 &\left.\, \times\left( \frac{Z^2}{M(u)^2} - \frac{K^2}{N(u)^2} - \epsilon \right) \right] = 0. \label{orbitleft}
\end{align}
As expected, the motion is only affected by the dislocation parameter $\chi$ if the momentum along the $z$-direction is considered.  Also, we have a couple between $\chi$ and the term due to the wiggles. If we only take the straight version of the string, $ \mathnormal{w}=0$ (absence of wiggles), the motion would be simplified, and the dislocation would not affect the particle's motion on the plane. 
If we make $\chi=0$ in Eq. \eqref{orbitleft}, we recover the geodesic equation of a wiggly cosmic string without dislocation \cite{dyda2007cosmic}\footnote{The result does not exactly match Eq. (13) in Ref. \cite{dyda2007cosmic} due to a difference in the sign of the third term. After redoing this calculation multiple times to check the sign, we are confident that the third term in our Eq. \eqref{orbitleft} is indeed negative.}. 
We also note that the massive and massless particles will have the exact geodesics in the case of null dislocation.

While complex, Eq. \eqref{orbitleft} when $\chi=0$ could be solved analytically to the first order in $ \mathnormal{w}$ using a perturbative approach. The steps of this solution method can be seen in Ref. \cite{dyda2007cosmic}, although they do not show plots with the geodesic curves. On the other hand, the complete form of Eq. \eqref{orbitleft} is significantly more complex due to the additional nonlinearity introduced by the $\chi$ terms and their nested dependencies, which complicate the perturbative expansion and subsequent solution steps. While an exact analytical solution might be theoretically possible, it will involve extensive algebraic manipulation that becomes impractical for manual solving. Therefore, a numerical approach is more suitable for this equation. Hence, we then provided specific numerical values for the involved parameters, constants of motion, and initial conditions for $u$. 

To obtain the geodesic curves along the $z$-direction (in three dimensions), we must obtain the geodesic equation for $z$. To find $z(\theta)$, we multiply $\dot{z}$ by $\dot{\theta}$. Both are found by solving the set of equations given in Eq. \eqref{CteMotion}, and we then get 
\begin{equation}
z^{\prime}-\frac{\alpha^2 Z }{u^2 {M(u)} \Omega(u) } + \frac{{\chi}}{\sqrt{M(u)}} = 0, \label{orbitz}
\end{equation}
where $z^{\prime}=\frac{dz}{d \theta}$. Here, as in \eqref{orbitleft}, a couple exists between the dislocation and wiggles. However, if we evaluate the limit of the absence of wiggles, the effect of $\chi$ will not disappear. 

The term $Z/\Omega(u)=Z/(J-\chi Z/\sqrt{M(u)})$ in Eqs. \eqref{orbitleft} and \eqref{orbitz} play an important role, as they can change the sign of the term it multiplies in the equation. Specifically, the equation changes signs if $J$ and $Z$ have different signs. Otherwise, the equation retains its original sign, whether they are both positive or negative. In the first case, the solution to Eq. \eqref{orbitz} is increasing (crescent, along $+z$-axis), while in the second case, the solution is decreasing (decrescent, along $-z$-axis). 
Additionally, we should note that our analyses avoid cases where $J=\chi Z/\sqrt{M(u)}$ to prevent divergences.
\begin{figure}[h]
\centering
\begin{subfigure}{0.1622\textwidth}
\includegraphics[width=\textwidth]{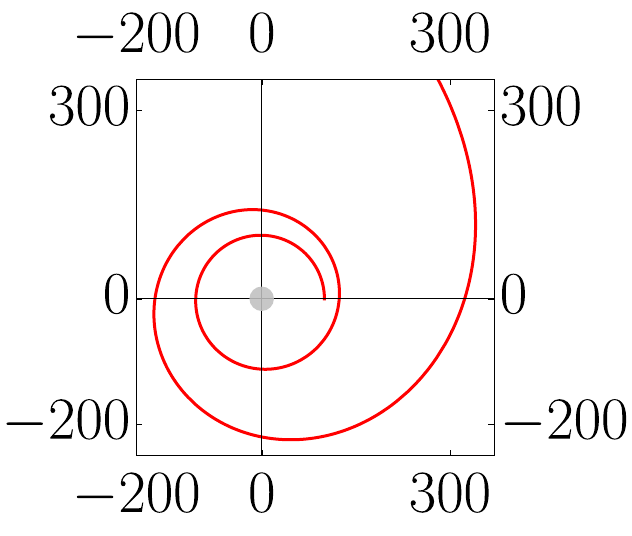}
\subcaption{$w=0$, $\chi=0$}
\end{subfigure}\hspace{1.2cm}
\begin{subfigure}{0.161\textwidth}
\includegraphics[width=\textwidth]{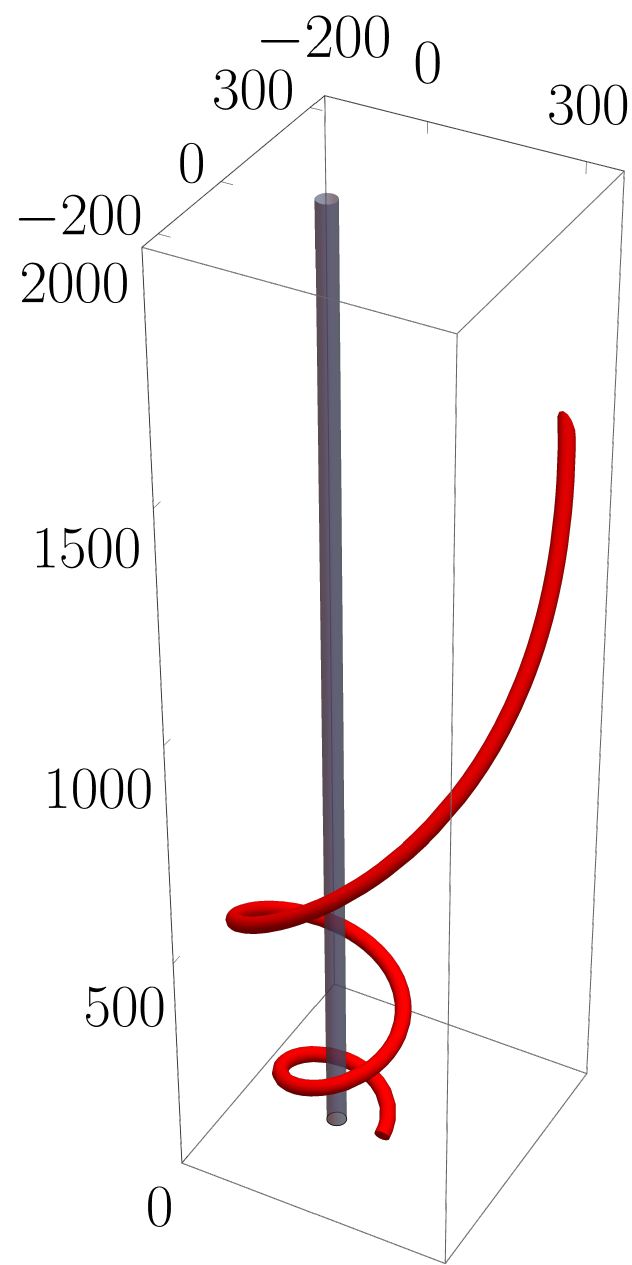}
\subcaption{$w=0$, $\chi=0$}
\end{subfigure}
\begin{subfigure}{0.1622\textwidth}
\includegraphics[width=\textwidth]{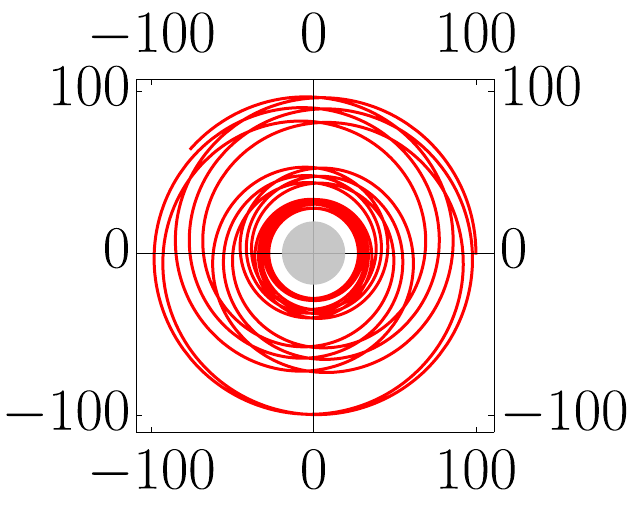}
\subcaption{$w=0.1$, $\chi=0$}
\end{subfigure}\hspace{1.2cm}
\begin{subfigure}{0.151\textwidth}
\includegraphics[width=\textwidth]{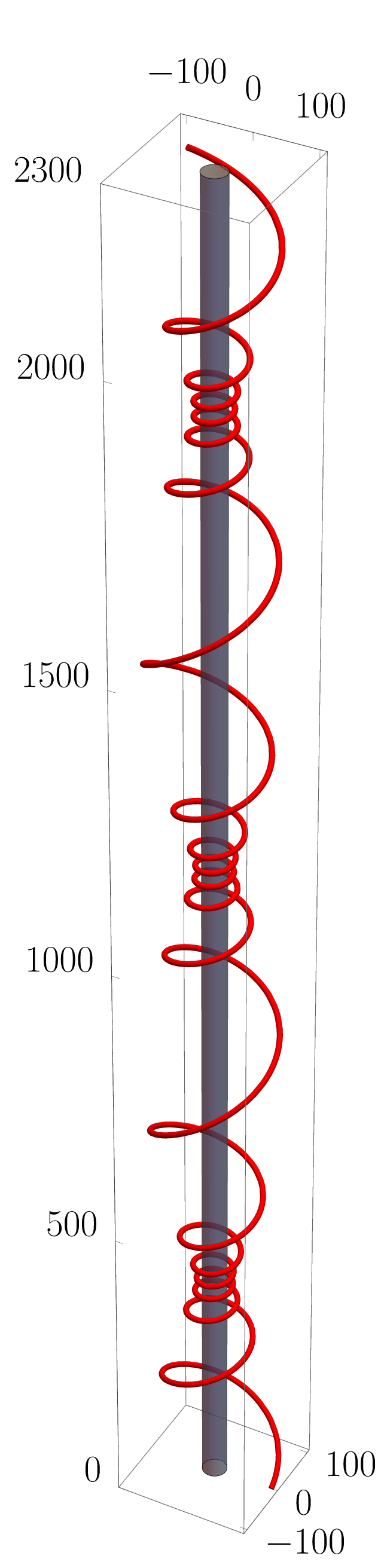}
\subcaption{$w=0.1$, $\chi=0$}
\end{subfigure}
\caption{\justifying Orbits without dislocation: Figures (a) and (b) show no wiggles, while Figs. (c) and (d) illustrate their effects. Parameters in all figures are fixed at  $Z=0.6$, $J=1.8$, $K=2$, and $\epsilon=-1$ (massive particles).}
\label{fig1}
\end{figure}
\begin{figure}[h]
\centering
\begin{subfigure}{0.161\textwidth}
\includegraphics[width=\textwidth]{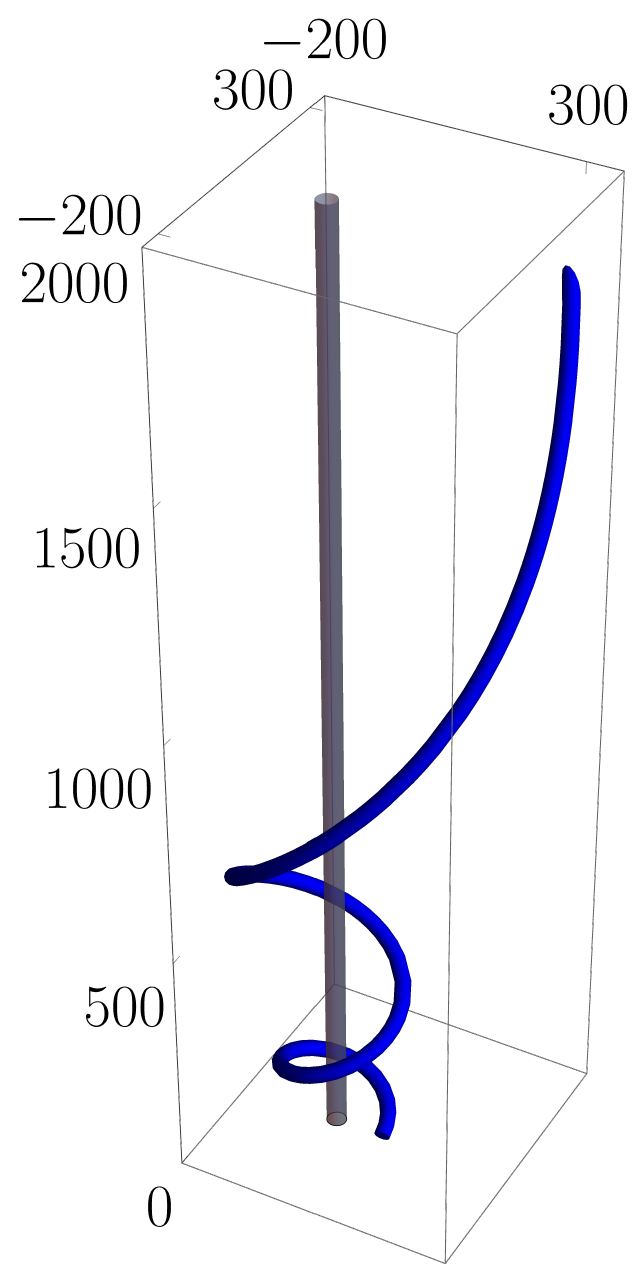}
\subcaption{$w=0$, $\chi=0.5$}
\end{subfigure}\hspace{0.0cm}
\begin{subfigure}{0.1622\textwidth}    \includegraphics[width=\textwidth]{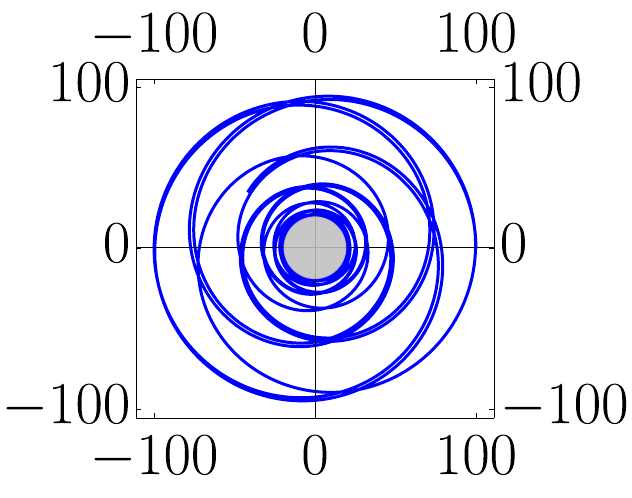}
\subcaption{$w=0.1$, $\chi=0.5$}
\end{subfigure}\hspace{0.0cm}
\begin{subfigure}{0.151\textwidth}
\includegraphics[width=\textwidth]{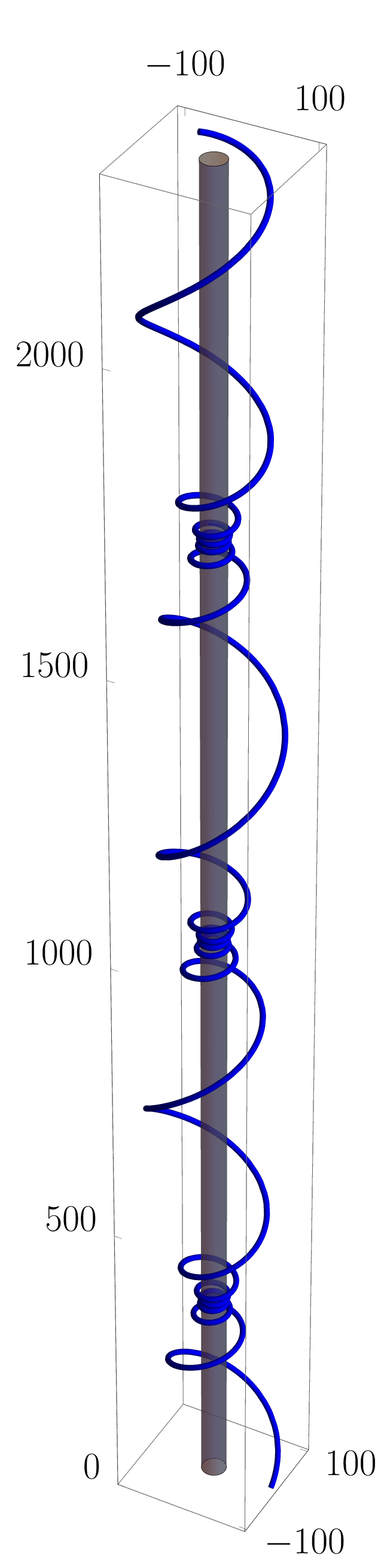}
\subcaption{$w=0.1$, $\chi=0.5$}
\end{subfigure}
\caption{Orbits under the effect of dislocation: Figure (a) shows the absence of wiggles but with dislocation, while Figs. (b) and (c) illustrates the presence of both wiggles and dislocation. We use the following parameters: $Z=0.6$, $J=1.8$, $K=2$, and $\epsilon=-1$.}
\label{fig2}
\end{figure}

To provide a clearer perspective on the points discussed above, we set up a numerical solution to Eqs. \eqref{orbitleft} and \eqref{orbitz} using mathematical software, demonstrating how the coupling between the dislocation and the wiggles affects the geodesics. The results, including the orbits found for specific constants of motion and parameters relevant to the physics of the problem, are presented and analyzed. We use a fixed small value of $\alpha=0.1$ to clearly illustrate the winding behavior of trajectories around the deficit angle of the string defect. Additionally, for the remainder of this paper, we adopt length units where $r_0=1$. Besides, the following plots take the initial conditions $u(0) = 0.01$, $u'(0) = 0$, and $z(0) = 0$, implying that the particles start their motion at $z = 0$ and $r = 1/u = 1/0.01 = 100$.

Figure \ref{fig1}(a) presents the geodesic projection of a moving particle on the plane, with $\mathnormal{w}=0$ and $\chi=0$. The orbits follow the conical geometry of the spacetime around the cosmic string; however, in the absence of wiggles, the orbits are unbounded. Figure \ref{fig1}(b) depicts the geodesics for a particle moving along the defect axis, also with $\mathnormal{w}=0$ and $\chi=0$. The motion starts at $u=100$ and only goes away from the defect, never returning to this value or another less than it.  In Fig. \ref{fig1}(c), the geodesic projection is shown for a moving particle on the plane with $\mathnormal{w} \neq 0$ and $\chi=0$. Here, we observe that due to the presence of wiggles, the orbits adhere to Bertrand's Theorem: they are bound, with motion between a minimum and maximum radius, without necessarily being closed. Figure \ref{fig1}(d) illustrates the geodesics for a particle moving along the defect axis with $\mathnormal{w} \neq 0$ and $\chi=0$. We observe that the particle encounters a barrier, preventing it from accessing the defect's core. Moreover, the particle remains radially bounded, explicitly showing the gravitational pull by the string. However, it can move freely along the $z$-axis, with its motion repeating in form with slight variations.
\begin{figure}[tbh!]
\centering
\begin{subfigure}{0.151\textwidth}
\includegraphics[width=\textwidth]{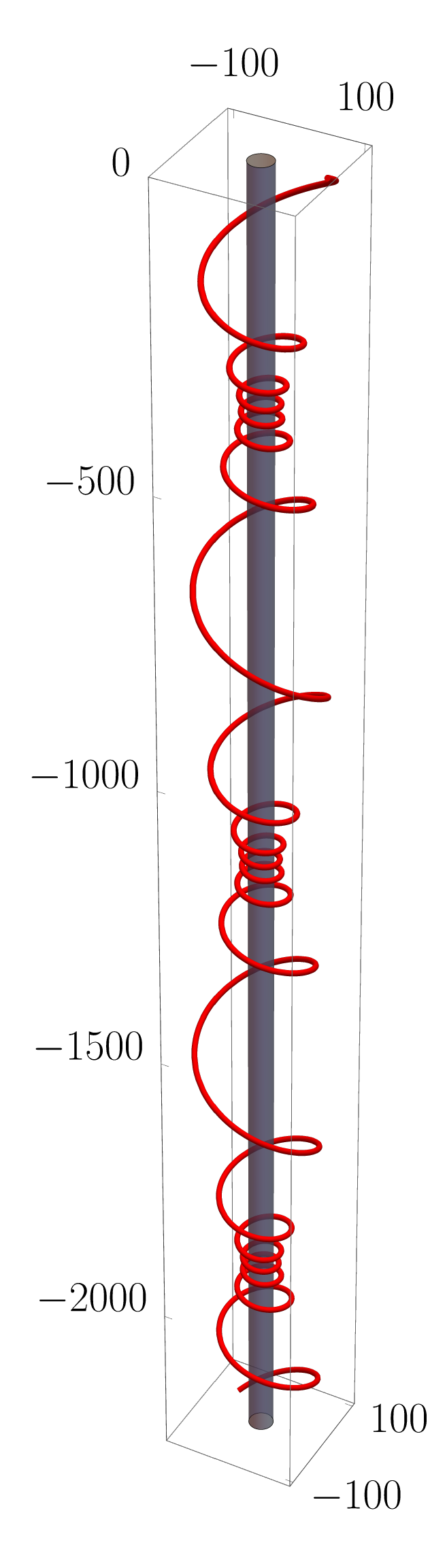}
\subcaption{$w=0.1$, $\chi=0$}
\end{subfigure}\hspace{0.0cm}
\begin{subfigure}{0.1622\textwidth}       \includegraphics[width=\textwidth]{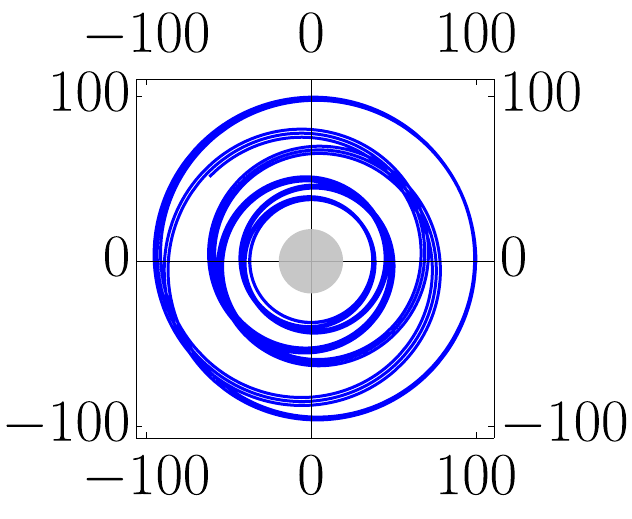}
\subcaption{$w=0.1$, $\chi=0.5$}
\end{subfigure}\hspace{0.0cm}
\begin{subfigure}{0.151\textwidth}
\includegraphics[width=\textwidth]{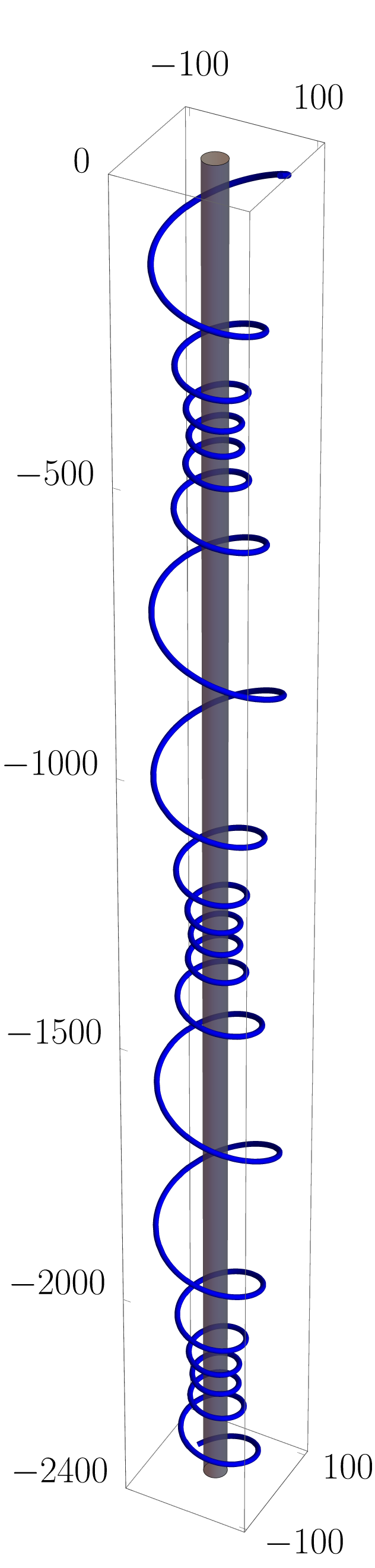}
\subcaption{$w=0.1$, $\chi=0.5$}
\end{subfigure}
\caption{(a) The same as in Fig. \ref{fig1}(d). (b) The same as in Fig. \ref{fig2}(b). (c) The same as in Fig. \ref{fig2}(c). However, we use the fixed values $Z=-0.6$, $J=1.8$, $K=2$, and $\epsilon=-1$.}
\label{fig3}
\end{figure} 

In Fig. \ref{fig2}(a), we observe the projection of geodesics for a moving particle on the plane with $\mathnormal{w}=0$ and $\chi\neq0$. Compared to Fig. \ref{fig1}(b), a noticeable curve stretching is evident due to dislocation. It is important to note that, according to Eq. \eqref{orbitleft}, when $\mathnormal{w}=0$, the dislocation parameter does not affect the motion projection on the plane and only influences the particle when it travels along the defect ($z$-axis).
Figure \ref{fig2}(b) illustrates that the radial barrier shifts toward the core of the defect when the dislocation is present, in contrast to Fig. \ref{fig1}(c), which features a larger inner region free of particle paths. Additionally, the distribution of paths in Fig. \ref{fig1}(c) appears more well-defined and less chaotic. Furthermore, the presence of dislocation in the three-dimensional space results in more compact helices, as highlighted in Fig. \ref{fig2}(c). Moreover, the helices are displaced to lower values along the $z$-axis.
\begin{figure*}[tbh]
\centering
\begin{subfigure}{0.1622\textwidth}
\includegraphics[width=\textwidth]{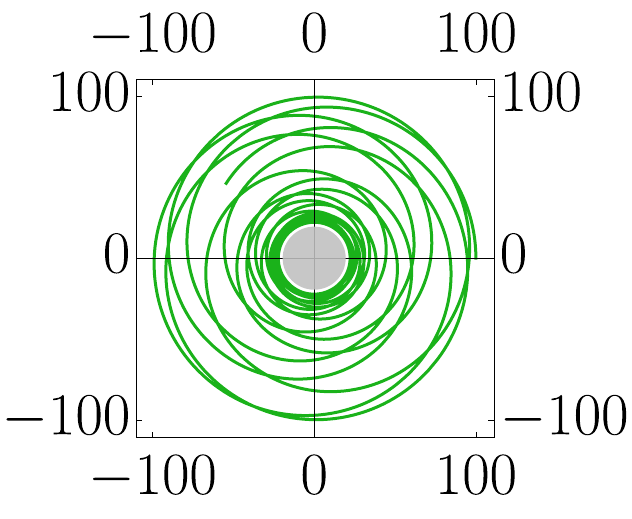}
\subcaption{$w=0.1$, $\chi=0.5$}
\end{subfigure}\hspace{1cm}
\begin{subfigure}{0.1515\textwidth}      \includegraphics[width=\textwidth]{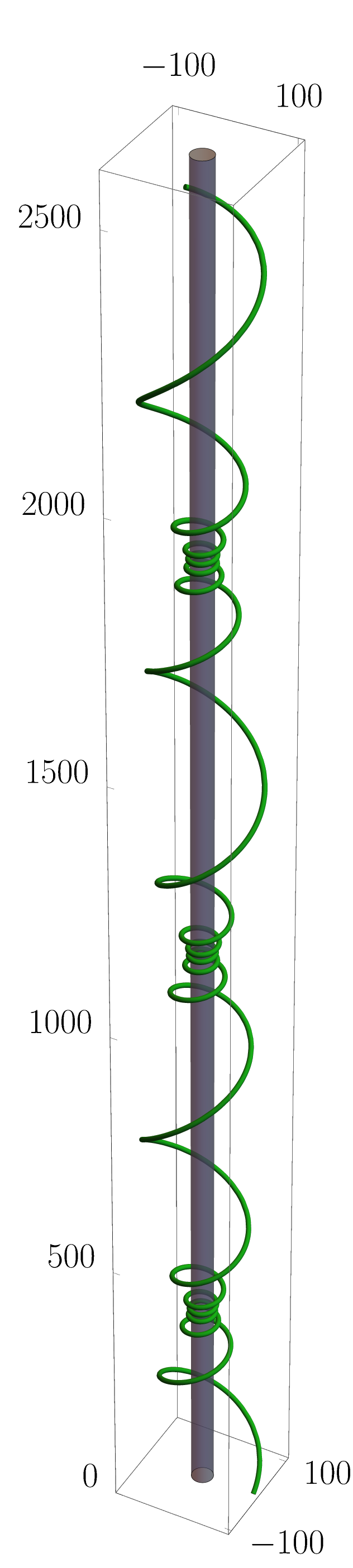}
\subcaption{$w=0.1$, $\chi=0.5$}
\end{subfigure}\hspace{2cm}
\begin{subfigure}{0.1622\textwidth}
\includegraphics[width=\textwidth]{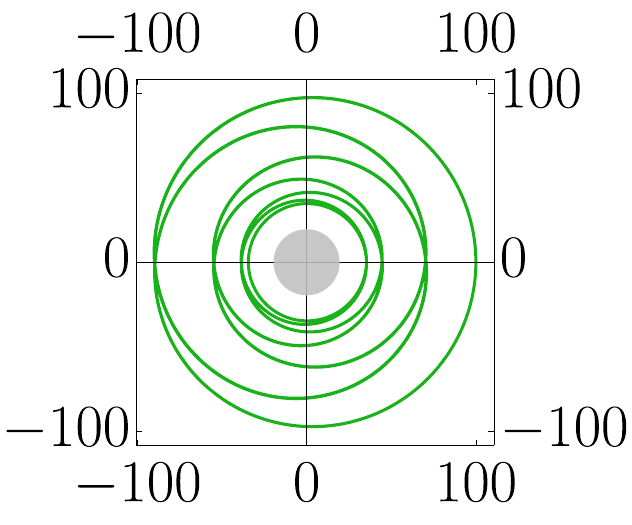}
\subcaption{$w=0.1$, $\chi=0.5$}
\end{subfigure}\hspace{1cm}
\begin{subfigure}{0.1515\textwidth}
\includegraphics[width=\textwidth]{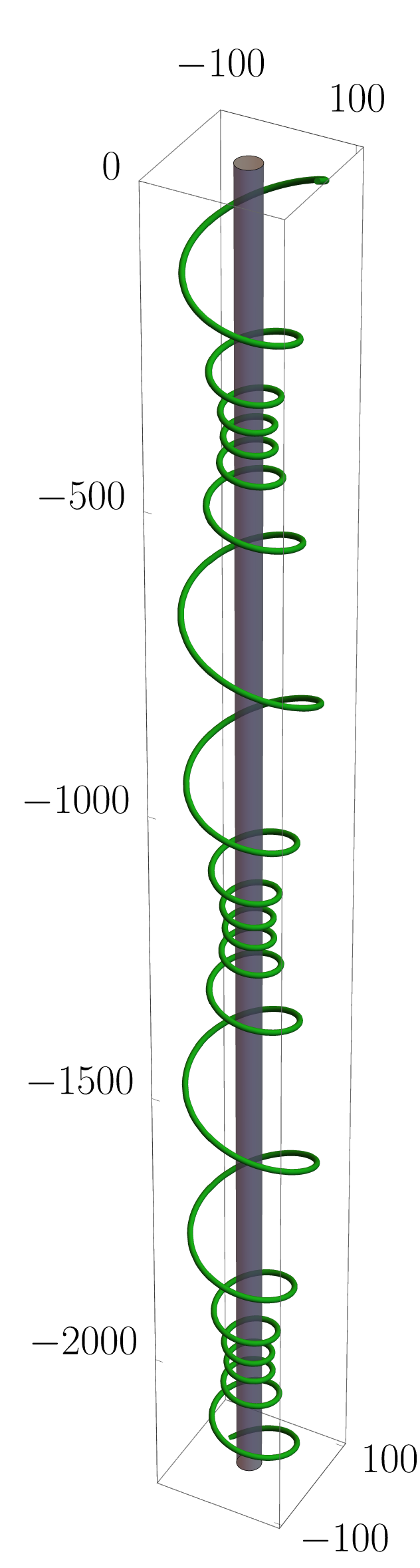}
\subcaption{$w=0.1$, $\chi=0.5$}
\end{subfigure}
\caption{(a) The same as in Fig. \ref{fig2}(b). (b) The same as in Fig. \ref{fig2}(c). (c) The same as in Fig. \ref{fig3}(b). (d) The same as in Fig. \ref{fig3}(c). However, here, we use $\epsilon=0$ (massless particles).}
\label{fig4}
\end{figure*}

In Fig. \ref{fig3}, we examine the case where the motion constants $J$ and $Z$ have opposite signs. Initially, we observe that the motion occurs along the negative $z$-axis, with the dislocation effect being notably different from when the particle moves along the positive $z$-axis. Comparing Fig. \ref{fig3}(a) with Fig. \ref{fig1}(d), we find that the effect is very similar, most differing only in the direction of motion. However, comparing Fig. \ref{fig3}(c) with Fig. \ref{fig2}(c) reveals a significant stretching of the helices, along with a displacement to lower values along the $z$-axis (this behavior can also be observed by comparing Figs. \ref{fig3}(a) and \ref{fig3}(c). Additionally, the inner region free of particle paths is now considerably larger. This is more clearly visible in Fig. \ref{fig3}(b).

Figure \ref{fig4} illustrates the behavior of massless particles. Comparing Fig. \ref{fig4}(a) with Fig. \ref{fig2}(b), we observe distinctly different orbits, particularly noting that massless particle orbits have a slightly larger inner center compared to those of massive particles. This difference appears to stem from the attraction experienced by massive particles due to the rod-like potential introduced by the wiggles.
Examining Fig. \ref{fig4}(c) and Fig. \ref{fig3}(b), specifically for motion along the negative $z$-axis, we observe a significant enlargement in the radius of the inner free region for massless particles. This is due to the fact that the centrifugal effect plays a more pronounced role on massless particles than on particles with mass. Furthermore, comparing Fig. \ref{fig4}(b) with Fig. \ref{fig2}(c), we note elongation of the helices and a displacement towards higher values along the $z$-axis.
In contrast, Fig. \ref{fig4}(d) compared to Fig. \ref{fig3}(c) shows the opposite trend: a contraction of the massless particle curves and a shift towards occupying a smaller range along the $z$-axis.
\begin{figure*}[tbh]
\centering
\begin{subfigure}{0.161\textwidth}
\includegraphics[width=\textwidth]{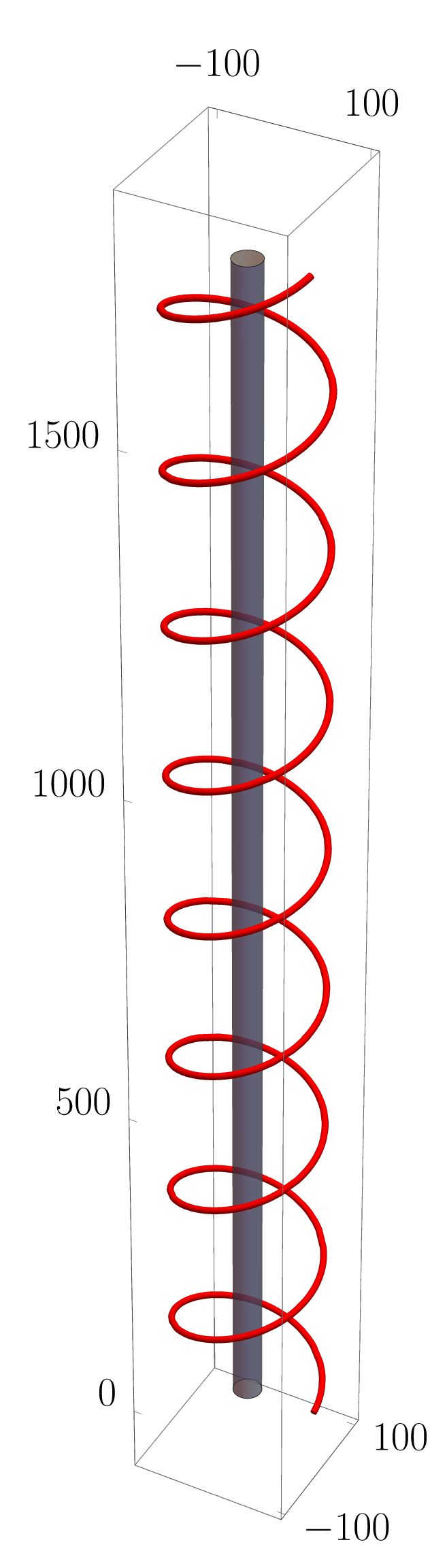}
\subcaption{$w=0.01$, $\chi=0$}
\end{subfigure}\hspace{1.cm}
\begin{subfigure}{0.161\textwidth}
\includegraphics[width=\textwidth]{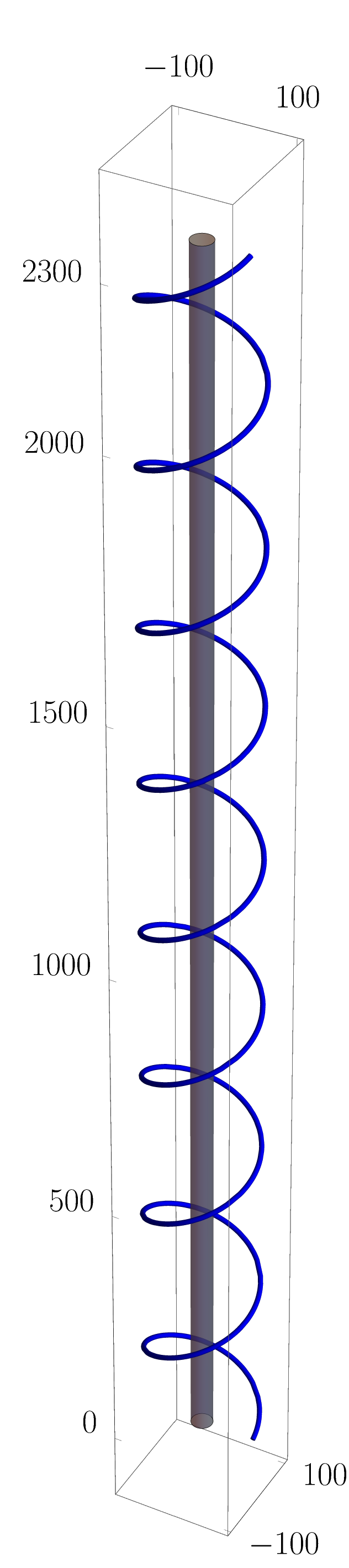}
\subcaption{$w=0.01$, $\chi=0.8$}
\end{subfigure}\hspace{2cm}
\begin{subfigure}{0.161\textwidth}
\includegraphics[width=\textwidth]{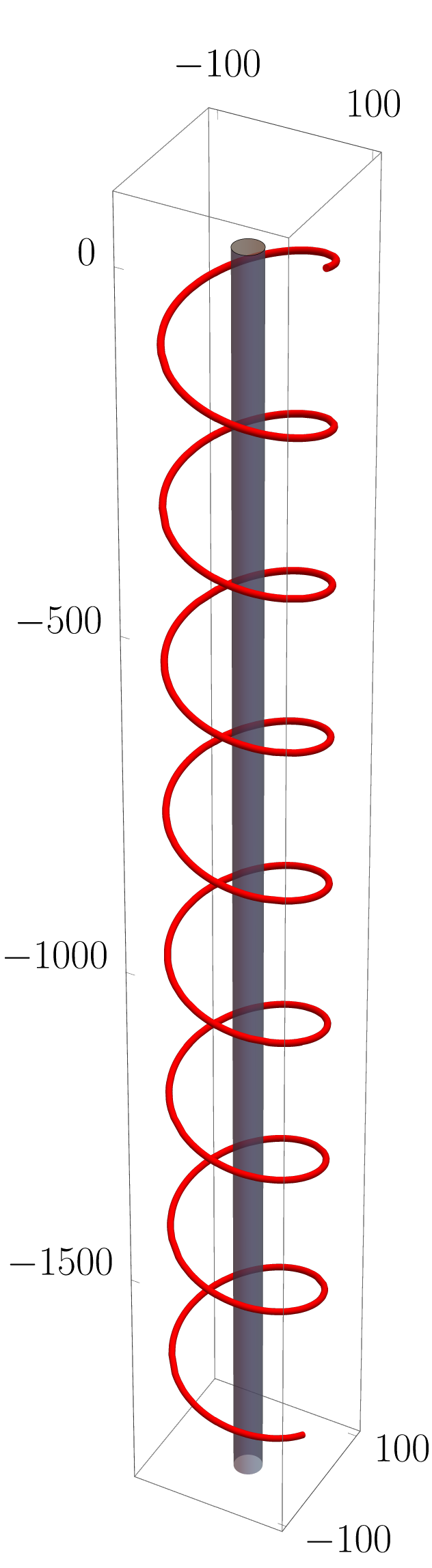}
\subcaption{$w=0.01$, $\chi=0$}
\end{subfigure}\hspace{1.cm}
\begin{subfigure}{0.161\textwidth}
\includegraphics[width=\textwidth]{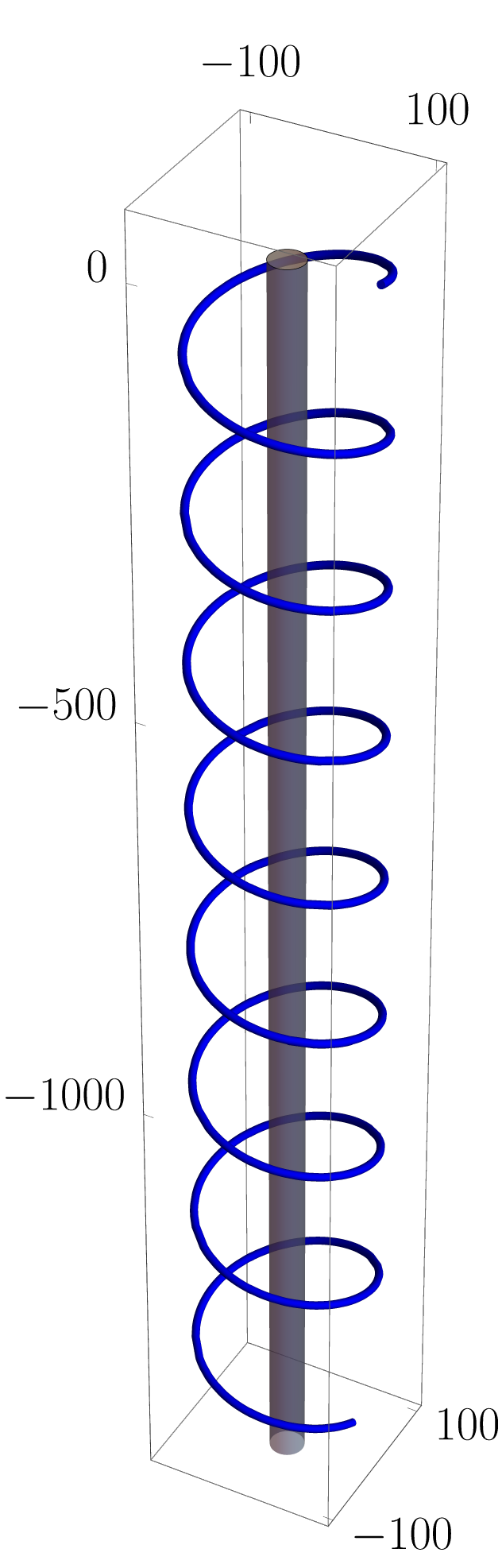}
\subcaption{$w=0.01$, $\chi=0.8$}
\end{subfigure}
\caption{Trivial orbits without dislocation (a) and with dislocation (b) for fixed values $Z=0.6$ and $J=1.8$. Figures (c) and (d) display the same comparison but for $Z=-0.6$ and $J=1.8$.}
\label{Fig5}
\end{figure*}
%\FloatBarrier

As mentioned earlier, the trivial solution $u' = 0$ leads to a circular orbit for the radial motion. However, the three-dimensional plot includes the solution for $z(\theta)$, which depends on the dislocation parameter, as demonstrated in Fig. \ref{Fig5}. Comparing Fig. \ref{Fig5}(b) with Fig. \ref{Fig5}(a), we observe a stretching effect when dislocation is introduced. Conversely, comparing Fig. \ref{Fig5}(d) with Fig. \ref{Fig5}(c), we observe a contraction in the presence of dislocation.
    
To conclude this section, we would like to highlight the most important result of this section: we observe that the presence of wiggles  ($\mathnormal{w}\neq 0$) not only makes bound orbits possible but also allows the radial orbits to be affected by $\chi$. The radial motion, described by Eq. \eqref{orbitleft}, is influenced by dislocation only when there is a coupling between $\mathnormal{w}$ and $\chi$ while the particle remains free to move along the $z$-axis. Without wiggles, the radial orbits do not depend on the dislocation parameter. However, Eq. \eqref{orbitz} shows that dislocation influences motion along the $z$-axis even without wiggles on the string.
Our results align with the geodesics found around line defects in elastic solids, as discussed in Ref. \cite{DEPADUA1998153}, where the authors demonstrate stretching the orbits when the Burgers vector is introduced. In their case, the metric does not consider the wiggles factor $\mathnormal{w}$. Additionally, Ref. \cite{azevedo2017wiggly} studies wiggly cosmic strings in the absence of dislocation using Eq. \ref{metric} and predicts two types of orbits from the effective potential of the Klein-Gordon (KG) equation for this geometry: radially bounded helices around the string with a constant radius, and radially bounded helices around the string with a minimum and maximum radius. These are exactly the two types of orbits we found here once the wiggles on the string are turned on. Moreover, we have observed that introducing the dislocation parameter and the subsequent coupling of it with the wiggly parameter results in a richer variety of orbit configurations.

\section{Relativistic quantum mechanics in the spacetime of a Wiggly Cosmic Dislocation}

In this section, we focus on solving the KG equation in the spacetime of a wiggly cosmic string with dislocation. This unique spacetime configuration, characterized by the coupling between small-scale structures and dislocation, introduces novel challenges and features, offering a rich context for exploring the behavior of scalar fields.
The KG equation has broad applications across multiple fields \cite{berestetskii1982quantum,greiner2000relativistic}.

The general form of the KG equation (with $\hbar=c=1$) in curved spacetime is given by
\begin{equation}
\frac{1}{\sqrt{-g}} \partial_\mu \left( \sqrt{-g} g^{\mu \nu} \partial_\nu \right) \psi - M^{2} \psi - \xi R \psi = 0,\label{kgc}
\end{equation}
where $M$ is the mass of the particle, $R$ is the Ricci scalar curvature, $\xi$ is a dimensionless coupling constant, and $\psi$ is the wave function. Including the Ricci scalar term enables the interaction between the scalar field and the curvature of spacetime, which is particularly significant in regions with strong gravitational fields. Specifically, the spacetime metric in Eq. \eqref{metric2} represents a linearized gravity solution (a weak-field approximation) with metric coefficients proportional to  $\mathnormal{w}$. After extensive manipulation involving the Riemann curvature tensor and Christoffel symbols, we find this metric's Ricci scalar proportional to $\mathnormal{w}^2$. Therefore, this small contribution can be neglected without any significant impact.

To derive the radial equation from \eqref{kgc}, we use the ansatz
\begin{equation}
\psi \left( t,r,\varphi ,z\right) = e^{-iEt}e^{im\varphi}e^{ikz}f(r),
\label{sol}
\end{equation}
where $E, k \in \mathbb{R}$ and $m=0,\pm1,\pm2,\cdots$. 
By substituting this ansatz into the KG equation and performing a detailed calculation, we obtain 
\begin{align}
&-\frac{1}{r}\frac{d}{dr}\left(r\frac{ df}{dr}\right)-(E^2 - k^2 - M^2) f+\frac{(m - k \chi)^2}{\alpha^2 r^2} f \notag \\
& + \left( 1 - \frac{  m k \chi - k^2 \chi^2}{\alpha^2 r^2} \right) \mathnormal{w}(E^2+k^2)\ln\frac{r}{r_0}f=0,
\label{eqradial0}
\end{align}
The spatial dislocation introduces a correction to the particle's angular momentum $m-k\chi$, similar to the effect observed in the motion of a charged particle orbiting around a tube with magnetic flux \cite{PhysRev.115.485,PhysRevLett.48.1144}, as noted in Refs. \cite{10.1063/1.531995,C.Furtado_2000,doi:10.1142/S0219887823500676,EPJP.2019.134.131}. In addition, the coupling between the wiggles and dislocation, represented by the term $k\chi\mathnormal{w}$, introduces an additional term $\propto (1/r^2) \ln r$. This term modifies the wave equation and leads to new effects on wave propagation. For instance, the product $mk\chi$ (in the fourth term of Eq. \eqref{eqradial0}), which only exists when considering particle propagation in the spacetime of a string with wiggles, can be either positive or negative because both $m$ and $k$ can take on positive or negative values. In short, we observe that beyond the expected change in angular momentum (as indicated in the third term of Eq. \eqref{eqradial0}, even in the absence of small-scale structures), dislocation induces novel and intriguing effects on wave propagation when combined with a wiggly cosmic string. This conclusion is supported by the results presented in Section \ref{section1}, where the coupling of wiggles with dislocation produces notable effects on the trajectories of particles. From this point onward, we will delve deeper into the effects of this coupling on wave propagation.        

By making a suitable variable
transformation $f(r) = r^{-1/2}R(r)$
and if we also consider the new radial variable as $r \to (E^2 + k^2)^{1/2}r$ and $r_0 \to (E^2 + k^2)^{1/2} r_0$,
we find the following radial equation in the eigenvalue form
\begin{equation}
-\frac{d^2 R(r)}{dr^2} + V_{eff}(r)R(r)= \zeta_{n,m} R(r),\; \text{with}\; n\in \mathbb{N},
\label{eqradial}
\end{equation}
where
\begin{equation}
V_{eff}(r)=  \frac{(m - k \chi)^2 - \frac{1}{4}}{\alpha^2 r^2} + \left( 1 - \frac{ m k \chi  - k^2 \chi^2}{\alpha^2 r^2} \right)\mathnormal{w} \ln\frac{r}{r_0}\label{veff}
\end{equation}
is the effective potential and 
\begin{equation}
\zeta_{n,m} =\frac{E^2 - k^2 - M^2}{E^2 + k^2}
\label{zeta_mn}
\end{equation}
are the eigenvalues. For each eigenvalue $\zeta_{n,m}$, we have associated  energies $E_{n,m}$ in Eq. (\ref{zeta_mn}) expressed as \begin{equation} E_{n,m}=\pm\sqrt{\frac{(1+\zeta_{n,m})k^2+M^2}{1-\zeta_{n,m}}}. \label{energy} \end{equation} 
Therefore, for each value of $\chi$, a different value of $\zeta_{n,m}$ is determined, consequently leading to a different value of $E_{n,m}$. It is also observed that the eigenvalues cannot exceed 1 ($\zeta_{n,m} < 1$), as the denominator in Eq. \eqref{energy} would otherwise result in imaginary energy. Hence, a constraint is established that limits the number of wave modes propagating along the wiggly string, which depends on the parameters of the string itself and the dislocation. In addition, we observe that introducing the mass term $M$ leads to a shift in the energy spectrum. By the way, only particles with energy $E_{n,m}$ greater than the shift term $M^2/(1-\zeta_{n,m})$ will propagate. Otherwise, the wave mode for the particle will become evanescent. Naturally, setting $M=0$ in this section makes all the results applicable to the propagation of massless particles, eliminating evanescent waves caused by the mass-dependent energy shift. However, the constraint from the denominator in the energy equation remains valid.  

Equation \eqref{eqradial} recovers Eq. (7) from Ref. \cite{azevedo2017wiggly}, where the influence of small-scale structures on field propagation was studied in the absence of dislocation. However, the logarithmic potential in Eq. \eqref{eqradial} lacks a constant coefficient due to the presence of dislocation (also depends on the parameter $k$). This likely causes significant changes in the wave functions, probability densities, and energies of the particle's bound states. Additionally, in the absence of both wiggles and dislocation, Eq. \eqref{eqradial} reduces to a Bessel-type differential equation, as expected for a system with cylindrical symmetry \cite{abramowitz1965handbook}.

The effective potential $V_{eff}(r)$, Eq. \eqref{veff}, can be analyzed by examining its two main terms. Namely, the first term is a centrifugal-like potential,
\begin{equation}
V_{\text{centrifugal}}(r) = \frac{(m - k \chi)^2 - \frac{1}{4}}{\alpha^2 r^2},
\label{eq:centrifugal}
\end{equation}
which is typical in radial problems and is associated with the particle's angular momentum. This term diverges as $r$ approaches zero, creating a strong repulsive barrier when $m - k \chi \neq 0$. The magnitude of this term is modulated by the parameters $m$, $k$, $\chi$, and $\alpha$, and for large values of $m - k\chi$, the barrier can become significant, preventing the particle from reaching the origin. 
The second term is a modified logarithmic potential
\begin{equation}
V_{\text{log}}(r) = \left( 1 - \frac{ m k \chi  - k^2 \chi^2}{\alpha^2 r^2} \right)\mathnormal{w} \ln\frac{r}{r_0}.
\label{eq:logarithmic}
\end{equation}
This term is more complex due to the combination of a logarithmic potential $\mathnormal{w} \ln\left(r/r_0\right)$ and a multiplicative factor that includes a correction depending on $1/r^2$. For large $r \to \infty$, the $1/r^2$ term becomes negligible, and the logarithmic potential dominates, resulting in a potential that increases slowly as $\ln(r)$. This suggests that the effective potential grows gently at large distances, indicating the possibility of bound states with discrete energy levels.
In contrast, for $r \to 0$, the $1/r^2$ correction can significantly alter the behavior of the potential. Depending on the values of $m$, $k$, and $\chi$, the factor $\left[ 1 - (m k \chi  - k^2 \chi^2)/\alpha^2 r^2 \right]$ may change the sign of the potential near the origin. If this factor is positive, the potential barrier near the origin is enhanced. In contrast, a negative factor could reduce the barrier or even create a potential well, modifying the structure of energy levels and the wave function. For small $r$, the centrifugal term in Eq.~\eqref{eq:centrifugal} dominates, creating a strong repulsive barrier, especially when $m - k \chi \neq 0$. This barrier influences the behavior of the wavefunction $R(r)$, leading to either oscillatory or exponentially decaying solutions near the origin. As $r$ increases, the logarithmic potential in Eq.~\eqref{eq:logarithmic} takes over, suggesting that the particle can be found in bound states with energies that increase slowly with distance. 
\begin{figure}[h!]
\centering	
\includegraphics[width=\columnwidth]{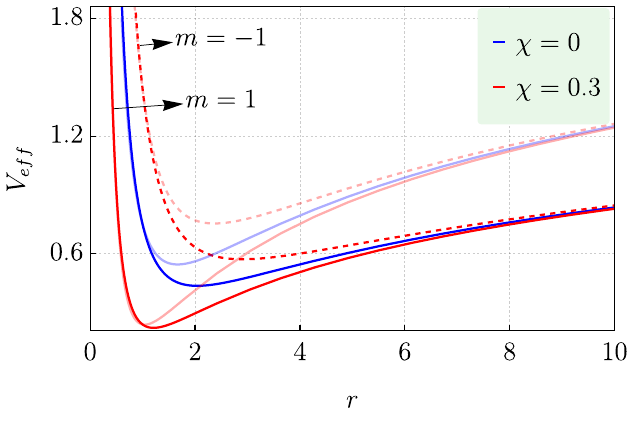}
\caption{Effective potential as a function of $r$ for different values of $\chi$. The solid blue line represents the profile with $m=\pm1, \chi=0$. We used the parameter values $\mathnormal{w} = 0.36$ and $k = 1$. 
The shaded lines represent $\mathnormal{w}=0.54$ (approximately $1.5$ times the value used for the other lines).} 
\label{fig:effective_potential_chi}
\end{figure}

\begin{figure}[h!]
\centering	
\includegraphics[width=\columnwidth]{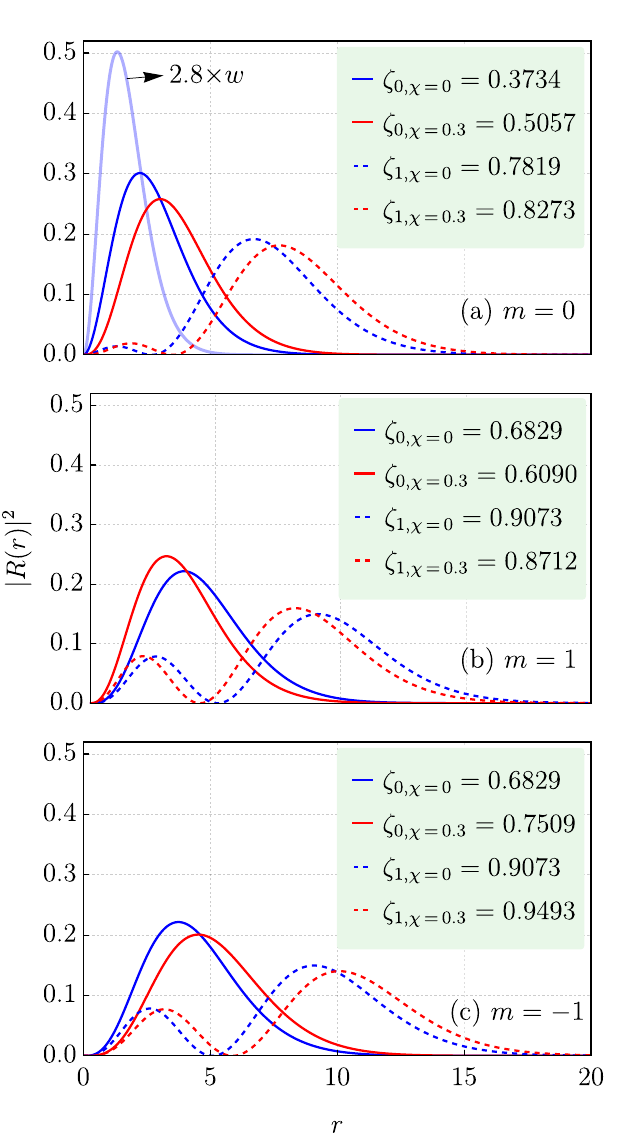}
\caption{Radial probability density function for different quantum numbers $n$ and $m$, with varying values of the parameter $\chi$. We used the parameter values $\mathnormal{w} = 0.36$ and $k = 1$. 
The shaded blue line represents $\mathnormal{w}=1$ (approximately $2.8$ times the value used for the other lines) and $\chi=0$ with the ground state given by $\zeta_{0}=0.5264$.}
\label{fig:probability}
\end{figure}

\begin{table}[h!]
\centering
\begin{tabular}{c@{\hspace{0.5cm}}c@{\hspace{0.25cm}}c@{\hspace{0.25cm}}c@{\hspace{0.5cm}}c@{\hspace{0.09cm}}}
\toprule
$m$ &  {$\zeta_{n, {m}}$ ($\chi=0$)} &   {$\zeta_{n, {m}}$ ($\chi=0.3$)} &  {$\zeta_{n, {m}}$ ($\chi=0.8$)} &$n$ \\
\midrule
%l0
&  0.3734 &  0.5057 & 0.6199 & 0 \\
  & 0.7819 &  0.8273& 0.8766 & 1 \\
0 & 0.9677 &  0.9949 & 1.0261 & 2 \\ 
& 1.0895 & 1.1089 & 1.1317 & 3 \\
  & 1.1802 & 1.1954& 1.2134 & 4 \\ \\
%l1
&  0.6829 & 0.6090 &0.4888& 0 \\
  &  0.9073 & 0.8712 &0.8212& 1 \\
1 &   1.0459  & 1.0224& 0.9914& 2 \\ 
& 1.1461 & 1.1289 & 1.1065& 3 \\
  & 1.2246  & 1.2110&1.1935& 4 \\ \\
%l-1 
& 0.6829 &  0.7509 & 0.8485 & 0 \\
 &0.9073 &  0.9493 & 1.0033 & 1 \\
-1 & 1.0459 &  1.0798 & 1.1126& 2 \\ 
& 1.1461 &  1.1647 & 1.1969 & 3 \\
  & 1.2246  & 1.2396 & 1.2662& 4 \\ \\
%l2
&  0.8479 &  0.7947 &0.6945 & 0 \\
  & 1.0032 & 0.9703 & 0.9142 & 1 \\
2 & 1.1127 & 1.0892 & 1.0508 & 2 \\ 
& 1.1971 &  1.1789& 1.1500 & 3 \\
  & 1.2666 & 1.2514& 1.2279 & 4 \\ \\
%l-2
&  0.8479 & 0.8962& 0.9673 & 0 \\
 & 1.0032 & 1.0349 & 1.0842 & 1 \\
-2  & 1.1127 &  1.1360 & 1.1734& 2 \\ 
& 1.1971 &  1.2154 & 1.2456 & 3 \\
  & 1.2666 & 1.2823& 1.3096 & 4 \\ \\
%l3
& 0.9607 & 0.9198 &0.8433 & 0 \\
 &1.0796 & 1.0510& 1.0005& 1 \\
3 & 1.1700 &  1.1482 & 1.1109& 2 \\ 
&  1.2429 & 1.2253 &1.1958 & 3 \\
  & 1.3074 & 1.2912& 1.2657& 4 \\ \\
%l-3
& 0.9607 &  0.9984 &1.0550& 0 \\
  &1.0796 & 1.1069 & 1.1493& 1 \\
-3 & 1.1700 &  1.1911 & 1.2248& 2 \\ 
&  1.2429 & 1.2603 & 1.2891 & 3 \\
  & 1.3074 &1.3241& 1.3534& 4 \\ 
  \toprule
\end{tabular} \\
\caption{Eigenvalues for different quantum numbers $n$ and $m$, with varying values of the parameter $\chi$. We used the parameter values $\mathnormal{w} = 0.36$ and $k = 1$.}
 \label{table1}
\end{table}

The parameter $\alpha$ plays a crucial role in determining the intensity of the $1/r^2$ terms, which are particularly important near $r = 0$. Smaller values of $\alpha$ increase the magnitude of these terms, leading to higher barriers and more tightly bound states. Meanwhile, the parameters $m$, $k$, and $\chi$ modulate both the centrifugal term and the correction to the logarithmic potential, affecting the number of nodes in the wave function and the overall shape of the potential. In this way, we can say that the effective potential $V_{\text{eff}}(r)$ exhibits a rich structure depending on the parameters $m$, $k$, $\chi$, and $\alpha$. 
Combining a centrifugal term and a modified logarithmic potential suggests the system may display significant potential barriers near the origin and bound states at large distances. The exact nature of the bound states and their energies is strongly influenced by the specific values of these parameters, with the possibility of complex and interesting behaviors for different ranges of $r$. 

In Fig. \ref{fig:effective_potential_chi}, we illustrate a particular profile of the effective potential, where we can see that when changing the configuration with  $m=1$ from $\chi=0$ to $\chi=0.3$ leads to a deeper and narrower potential well closer to the origin, which enhances the likelihood of forming bound states. Conversely, when $m=-1$, the effect is the opposite, resulting in a less deep and less narrow potential well further from the origin. The black line represents the case of $\chi=0$, indicating no change in the potential for quantum numbers $m=1$ and $m=-1$ (this can be easily understood by examining the effective potential).
\begin{figure*}[!thp]
\includegraphics[scale=1]{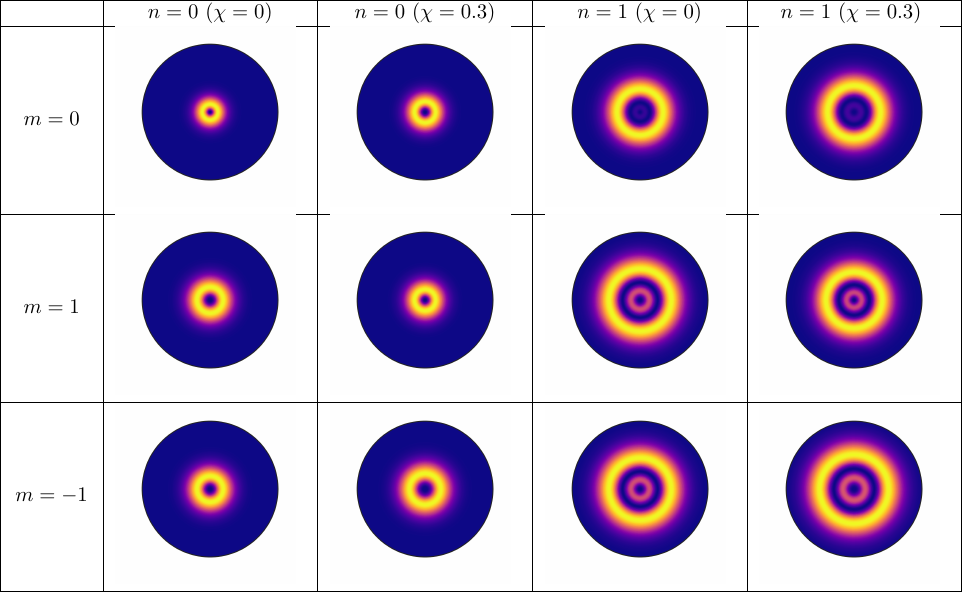} 
\caption{Disks of radial probability density highlighting regions of higher probability (higher intensity) for quantum numbers $n = 0, 1$ and $m = 0, 1, -1$, with varying dislocation parameter $\chi$. We used the same parameters as in Fig. \ref{fig:probability}.
}
\label{fig:intensity}
\end{figure*}
From Eq. \eqref{veff}, we observe that the difference for $\chi\neq0$ arises due to the presence of the $m k \chi$ term, which stems from the coupling between the wiggles and dislocation, as given in the metric in Eq. \eqref{metric2}. Notably, the plot for $m=-1$ is equivalent to the plot for $m=1$ with $k=-1$, a pattern that also holds for higher values of $m$ (e.g., $m=2$ and $m=-2$). As discussed in Section \ref{section1}, this indicates distinct effects depending on whether we travel along the positive $z$-axis or not.
The shaded lines highlight the effect of wiggles, parameterized by $\mathnormal{w}$. As $\mathnormal{w}$ increases, the potential well becomes less deep, indicating that the quantum states corresponding to wave functions allowed to propagate along the string are inversely proportional to the excess energy density, $\zeta_{n,m} \propto 1/\mathnormal{w}$ (this fact was also verified in Ref. \cite{azevedo2017wiggly}). Since for a GUT string $\mathnormal{w} \approx 10^{-6}$, the number of wave functions able to propagate along the string remains large.

It is important to mention that the effective potential reveals consistency with the previous section on geodesics. As we inspect the classical limit of the trajectories of particles from the effective potential given in Fig. \ref{fig:effective_potential_chi}, which only accommodates bound states, the trajectories are expected to follow the geometry of the system and to be radially bound helices around the defect. When the ``total energy'' $\zeta$, in Eq. \eqref{eqradial}, is at the minimum of the effective potential, the helices are expected to have a constant radius. When the energy is above this minimum, the radius varies between a smaller and a larger value determined by the potential. Although the trajectories follow this rule, their visual appearance differs mainly due to the influence of the dislocation parameter and its coupling terms with the wiggly parameter, as seen in Section \ref{section1}.  
\begin{figure}[h!]
\centering	
\includegraphics[scale=0.77]{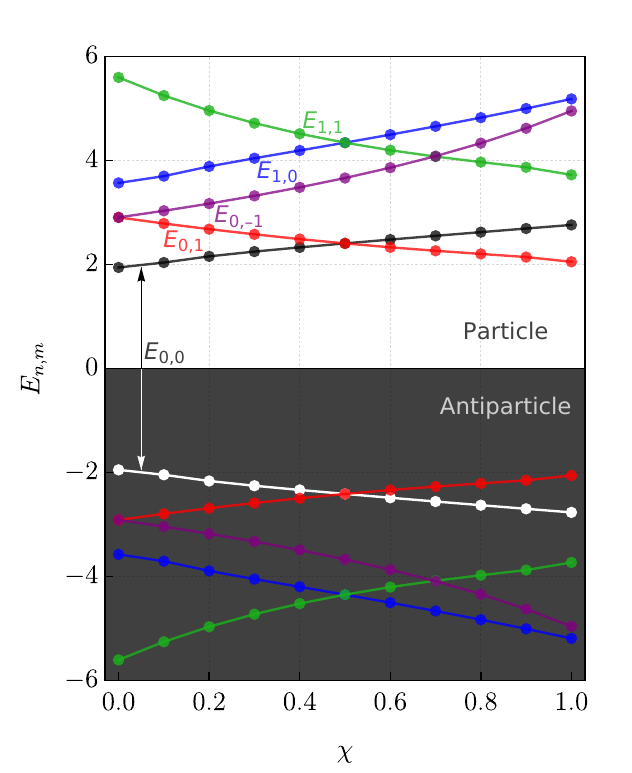}
\caption{The energy of the particle, $E_{n,m}$, as function of the dislocation parameter $\chi$. We used the parameter values $\mathnormal{w} = 0.36$ and $k = 1$. }
\label{fig:energy_versus_chi}
\end{figure}

To solve Eq. (\ref{eqradial}), we numerically compute the eigenvalues and corresponding eigenfunctions of the particle. The different states are labeled by the quantum numbers $n$ (the radial quantum number, starting from $0$, as we are in cylindrical coordinates) and  $m$ (the angular quantum number). Our approach is based on the Finite Difference Method \cite{burden2011numerical}, which approximates the differential operator using simple differences in line with the definition of a derivative. This method offers better stability characteristics than other methods for boundary-value problems. We use a fixed value of $\alpha = (1 - 10^{-6})^{1/2}$, which is commonly adopted in string theory \cite{MBHindmarsh_1995}. Therefore, the focus of this paper is on analyzing the effects of changes in the dislocation parameter $\chi$ and the wiggly parameter $\mathnormal{w}$.

Figure \ref{fig:probability} shows the radial probability density for different quantum numbers $n$ and $m$ for two values of $\chi$. When $m = 0$, the probability of finding the particle closer to the origin is more significant when the dislocation is absent ($\chi = 0$). However, for $m = 1$, the effect is reversed -- the probability increases with the presence of dislocation ($\chi \neq 0$). For $m = -1$, the particles are pushed away from the origin, and the probability amplitude decreases when dislocation is introduced.
It is important to note that, due to the term $mk\chi$ in Eq. \eqref{eqradial}, the solution for $m = -1$ is equivalent to the solution for $m = 1$ with $k = -1$, corresponding to a particle traveling in the opposite direction along the negative $z$-axis. In other words, without dislocation, increasing $m$ from 0 to 1 (or $m = -1$) creates a centrifugal effect. However, this effect no longer applies to particles traveling along the positive $z$-axis once dislocation is present. Instead, particles traveling in the opposite direction (negative $z$-axis) are pulled toward the defect's core.
This result aligns with Section \ref{section1}, where the term $Z/\Omega(u)$ shows that trajectories are closer to the defect for particles traveling along the positive $z$-axis and pushed away when traveling in the opposite direction. In summary, there is a centrifugal effect for negative $m = -1$, while for positive $m = 1$, particles are pulled toward the origin. This is evident as the eigenvalue decreases for $m = 1$ and increases for $m = -1$ when dislocation is present. The shaded line illustrates that a larger $\mathnormal{w}$ value results in a stronger bound state (curves pulled closer to the defect core). Notably, for $\mathnormal{w} = 1$ (with $\chi = 0$), the ground state eigenvalue is approximately 0.5264, which agrees with the results of Ref. \cite{azevedo2017wiggly} for field propagation in the metric given in Eq. \eqref{metric}, where dislocation is not considered. Additionally, this eigenvalue is equivalent to that found in the study of the influence of cosmic strings on a two-dimensional hydrogen atom \cite{DOSSAZEVEDO2024169660}, where the wave equation is mathematically analogous to Eq. \eqref{eqradial}, without dislocation.

Figure \ref{fig:intensity} presents the same effects as Fig. \ref{fig:probability}, but in a more visually appealing manner. The changes in the size of the rings, corresponding to variations in the quantum numbers and the presence of dislocation, are clearly visible and align with the earlier discussion of Fig. \ref{fig:probability}.

Table \ref{table1} presents the eigenvalues for higher quantum numbers and dislocation parameters, compared to those shown in Fig. \ref{fig:probability}. Some of the quantum states listed in this table are further illustrated in Fig. \ref{fig:energy_versus_chi}. The table also shows that some of the eigenvalues exceed $1$, which was expected given our choice of $\mathnormal{w}=0.36$. While this value is exaggeratedly large for a GUT string, it is an appropriate choice for comparison with the $\mathnormal{w}=1$ value in Fig. \ref{fig:probability}.

Figure \ref{fig:energy_versus_chi}
 depicts the particle energy levels as a function of $\chi$, based on the eigenvalues shown in Fig. \ref{fig:probability}
 and updated according to Eq. \eqref{energy}. As mentioned earlier, the energy levels increase with dislocation for the states with $m=0$ and $m=-1$. In contrast, the energy decreases with dislocation for the states with $m=1$.
It is important to note that this change occurs in different ways. For instance, the increase is nearly linear for the state ($n=0,m=0$). A similar trend is observed for the state $(n=1,m=0)$, but with faster growth. In contrast, the state $(n=0,m=-1)$ exhibits parabolic-like behavior. For values of $\chi \geq 0.5$, some states show degeneracy, or states with lower eigenvalues can surpass those with higher eigenvalues as $\chi$ increases, and vice versa. This behavior is particularly evident for the states $(n=0, m=0)$ and $(n=1, m=0)$, which show significant changes at $\chi = 0.5$. This interesting behavior will lead to an opposite trend in the radial probability density compared to that shown in Figs. \ref{fig:probability} and \ref{fig:intensity}, as the value of $\chi$ exceeds $0.5$. For comparing these states and others, see Table \ref{table1}.
Lastly, we see that for each positive particle energy, there is a corresponding symmetric negative antiparticle energy, as expected from a KG equation solution.

Before ending this section, it is important to emphasize that the novel effects described here, such as the differences in solutions for waves traveling along the positive and negative $z$-axis, only occur when both wiggles and dislocation are present.

\section{\label{sec:concl}Conclusions}

In this research, we investigated the spacetime of a wiggly cosmic dislocation by analyzing the effects of the coupling between small-scale structures (wiggles) and dislocation on the dynamics of massive and massless particles. By studying both the geodesic motion and the propagation of scalar fields through the Klein-Gordon equation, we have demonstrated how these two factors, when combined, produce novel and significant changes in particle behavior.

From the geodesic analysis, we observed that while wiggles alone result in bound orbits for particles, the presence of dislocation introduces a coupling between the angular and linear momentum components, particularly affecting motion along the $z$-axis. The combination of wiggles and dislocation reshapes the particle's trajectory, leading to more complex and confined paths compared to the simpler orbits found in straight cosmic strings. The dislocation parameter, $\chi$, plays a critical role in modifying these trajectories, causing a stretching or contracting effect depending on whether the particle moves along the negative or positive $z$-axis. Moreover, we noted that dislocation alone does not influence the radial motion without wiggles but still affects the particle's trajectory along the $z$-axis.

The study of the KG equation further revealed that the coupling between wiggles and dislocation significantly modifies the effective potential felt by particles. We identified the formation of potential wells, whose depth and structure depend on the string parameters. This leads to the formation of bound states with discrete energy levels. Our numerical results showed that the wiggly parameter $\mathnormal{w}$ and the dislocation parameter $\chi$ alter the system's energy spectrum, wave functions, and probability densities. Notably, we found that for quantum states with $m=1$, increasing the dislocation parameter decreases the energy levels, while for states with $m=-1$ or $m=0$, the energy levels increase with dislocation. This asymmetry highlights the intricate role of dislocation in wave propagation.  It is worth noting that, due to the term $mk\chi$, a state with a negative value of $m$ has an energy level equivalent to that of a state with the symmetric positive $m$ but with a negative wavenumber $k$ (indicating a wave traveling along the negative $z$-axis). Thus, the direction of propagation plays a crucial role in the outcome of the solution.

Overall, the combined effects of wiggles and dislocation in cosmic string spacetime reveal a rich structure of particle behavior with similar and consistent implications for both particle motion and wave propagation. These results contribute to a deeper understanding of the role of small-scale structures and dislocation in cosmic defects, offering new theoretical insights that may be relevant to cosmological models and further explorations in high-energy astrophysical phenomena.

Finally, we will continue using the numerical techniques employed in this study and plan to extend this work by investigating the influence of magnetic fields on charged particle motion and field propagation in the spacetime of a wiggly cosmic dislocation. Any further advancements will be reported in future publications.

\section*{\label{sec:acknw}Acknowledgement}

This study was partially supported by the Brazilian agencies CAPES—Finance Code 001, CNPq, and FAPEMA. F. S. Azevedo acknowledges CNPq Grant No. 153635/2024-0. Edilberto O. Silva acknowledges the support from the Grants CNPq/306308/2022-3, FAPEMA/UNIVERSAL-06395/22, and FAPEMA/APP-12256/22. This study was partly financed by the Coordenação de Aperfeiçoamento de Pessoal de N\'{\i}vel Superior - Brazil (CAPES) - Code 001.

\bibliographystyle{apsrev4-2}

\end{document}